\documentstyle[amssymb,aps,prb]{revtex}

\begin{document}
\title{Photoinduced IR absorption in (La$_{1-x}$Sr$_{x}$Mn)$_{1-\delta }$O$_{3}$:
changes of the anti-Jahn-Teller polaron binding energy with doping}
\author{}
\author{T. Mertelj$^{1,2}$, D. Ku\v{s}\v{c}er$^{1}$, M. Kosec$^{1}$ and D. Mihailovic%
$^{1}$}
\address{$^{1}$Jozef Stefan Institute, P.O.Box 3000, 1001 Ljubljana, Slovenia}
\address{$^{2}$University of Ljubljana, Faculty of Mathematics and Physics,\\
Jadranska~19,1000 Ljubljana,Slovenia}
\maketitle

\begin{abstract}
Photoinduced IR absorption was measured in (La$_{1-x}$Sr$_{x}$Mn)$_{1-\delta
}$O$_{3}$. A midinfrared peak centered at $\sim 5000$ cm$^{-1}$
was\thinspace observed in the $x=0$ antiferromagnetic sample. The peak
diminishes and softens as hole doping is increased. The origin of the
photoinduced absorption peak is atributted to the photon assisted hopping of
anti-Jahn-Teller polarons formed by photoexcited charge carriers, whose
binding energy decreases with increasing hole doping. The shape of the peak
indicates that the polarons are small.
\end{abstract}

\section{Introduction}

Manganites with the chemical formula (R$_{1-x}$A$_{x}$)MnO$_{3}$ (R and A
are trivalent rare-earth and divalent alkaline-earth ions respectively) in
which colossal magnetoresistance is observed\cite
{KustersSingelton89,HelmoltWecker93} are similar to high $T_{c}$ cuprates in
many ways. In both families of compounds the physical properties are
strongly influenced by changing carrier concentration in relevant $d$ and$\
p $ derived electronic bands, and the interplay between lattice and
electronic degrees of freedom is essential for the low energy physics of both%
\cite{electronic-lattice}.

It has been shown in high $T_{c}$ cuprates that measurements of the
photoinduced (PI) absorption are a useful tool for investigating the nature
of low-lying electronic states \cite
{KimHeeger87,KimHeeger88,TalianiZamboni88,LiKim96}, especially in the range
of weak hole doping where PI absorption spectra are interpreted in terms of
the photon-assisted hopping of small-polarons\cite{MihailovicFoster90}.

On the other hand features in optical\cite
{KaplanQuijada96,OkimotoKatsufuji97,JungKim98,MachidaMoritomo98,MachidaMoritomo98clust,QuijadaCerne98,JungKim99}
and Raman scattering\cite{YoonLiu98} spectra of cubic/pseudo-cubic
manganites as well as related layered compounds\cite{CalvaniPaolone96}
indicate presence of small polarons. Since photoinduced transient changes of
physical properties have already been observed in this class of compounds%
\cite{MiyanoTanaka97,CoxRadaelli98,ZhaoShreekala98}, we expect that
photoinduced infrared absorption might reveal interesting spectral
information about polarons also in manganites.

In this paper we present PI absorption measurements in (La$_{1-x}$Sr$_{x}$Mn)%
$_{1-\delta }$O$_{3}$ at various hole doping, concentrating mostly on the
weak hole doping levels. We find clear evidence of a photoinduced
mid-infrared polaronic peak and we apply theoretical analysis to the spectra
in order to deduce the dependence of the polaron binding energy on doping.

\section{Experimental}

\subsection{Sample preparation and experimental setup}

The method of preparation and characterization of ceramic samples with
nominal composition (La$_{1-x}$Sr$_{x}$Mn)$_{1-\delta }$O$_{3}$ ($x=0$, 0.1,
0.2) has been published elsewhere\cite{HolcKuscer97}. Since cation
deficiency is common to this class of materials, we performed AC
susceptibility measurements on powders in the 50K$-$350K range to establish
the Mn$^{4+}$ content from their Curie temperatures. The the two Sr doped
samples which we used had $T_{C}\approx 210$K and $T_{C}\approx 340$K
(midpoint, see Fig. 1) for $x=0.1$ and $0.2$ respectively. It was found that
despite the absence of Sr the sample with $x=0$ showed a transition to the
ferromagnetic (FM) state at $T_{C}\approx 170$K. One portion of the $x=0$
sample was treated at 900${{}^{\circ }}$C for 300 min in Ar flow\cite
{HuangSantoro97} to decrease cation deficiency. This sample showed no sign
of any ferromagnetic transition and so it was concluded that $\delta $ is
sufficiently small that it is antiferromagnetic (AFM) and insulating below $%
T_{N}=140$K\cite{HuangSantoro97,UrushibaraMoritomo95}. According to the
phase diagram in Urushibara {\em et al.}\cite{UrushibaraMoritomo95} the FM $%
x=0$ and $x=0.1$\thinspace samples are insulating below $T_{C}$, while the $%
x=0.2$ sample is metallic below $T_{C}$ and shows giant magnetoresistance
(GMR) around $T_{C}$.

Normal and photoinduced transmittance measurements in the MIR and NIR
regions were performed by means of two separate Bomem MB series Fourier
transform spectrometers set at 16 cm$^{-1}$ resolution. The powder samples
were mixed with KBr powder in 0.1-0.2 wt. \% ratio and pressed into 12 mm
diameter pellets. The pellets were mounted in the Oxford liquid-He flow
optical cryostat equipped with KRS-5 windows. Special care was taken that
the KBr pallet was in a good thermal contact with the sample holder.

The excitation CW Ar$^{+}$-ion-laser light with 514.5 nm wavelength ($h\nu
=2.41$ eV) was guided into the cryostat by an optical fibre. Due to space
limitations in the cryostat the size of the spot illuminated by laser was
only $\sim 5$ mm in diameter in the center of the pellet. The maximum
excitation optical fluence $\Phi $ was $\sim $500 mW/cm$^{2}$. The laser
excitation was switched on and off by an electro-mechanical chopper placed
in the laser beam path.

To minimize heating effects due to laser light absorption in the sample and
instrumental drift, the PI spectra were taken by repeating one sample scan
(with excitation laser on) and one reference scan (laser off) approximately
every two seconds. At each temperature the thermal-difference (TD)
transmittance change was also measured without laser excitation by first
measuring the reference spectrum and then increasing the sample holder
temperature by 2K and measuring the sample spectrum. To remove any drifts,
the same procedure was then inverted to measure the sample spectrum 2K above
the given temperature first. The thermal difference (TD) transmittance
change was then obtained by averaging both spectra.

\subsection{Experimental results}

The transmittance spectra ${\cal T}$ of the samples are shown in Fig. 1.
First let us discuss the normal (non-photoinduced) infrared phonon spectral
bands. In all FM samples we observe two IR phonon bands at 640 cm$^{-1}$ and
400 cm$^{-1}$. There is no appreciable shift as a function of $x$ in the FM
region of doping. In the AFM sample, the lower 400-cm$^{-1}$ phonon band is
split in two bands at 376 cm$^{-1}$ and 420 cm$^{-1}$ which appear further
split at the peak. There are also two additional shoulders at 457 and 509 cm$%
^{-1}$. The appearance of the additional phonon bands in the AF phase is in
accordance with expected behaviour arising from its lower point symmetry due
to the static rotation and Jahn-Teller (JT) distortion of the MnO$_{6}$
octahedra compared to the FM phases.\cite{HuangSantoro97,ElemansVanLaar71}
The high frequency phonon band is shifted downwards to 585 cm$^{-1}$ with
respect to the FM samples and appears asymmetric with a shoulder like tail
extending toward high frequencies. It is not clear whether this is a single
phonon band or two overlapping bands.

In addition to the phonon bands we observe a broad absorption, which
increases in amplitude and shifts to lower energy forming a MIR peak as the
hole doping is increased in agreement with previously reported optical data. 
\cite{OkimotoKatsufuji97} (It should be noted that the shape of the peak in
the NIR region is strongly influenced by the scattering in the KBr pellet.
Fortunately, the spectral distortions due to the scattering are canceled out
in PI absorption spectra.)

The low temperature ($T=25$K) PI transmittance $(\frac{\Delta {\cal T}_{PI}}{%
{\cal T}})$ spectra of all four samples are shown in Fig. 2a. In addition,
the TD transmittance $(\frac{\Delta {\cal T}_{TD}}{{\cal T}})$ spectra taken
at the same temperature are shown in Fig. 2b. In the $x=0$ AFM sample a
strong broad PI midinfrared (MIR) absorption (negative PI transmittance)
centered at $\thicksim 5000$ cm$^{-1}$ ($\thicksim 0.62$ eV) is observed.
The virtually flat TD spectrum in Fig. 2b shows that the PI absorption is
clearly not due to laser heating effects. In the frequency range of the
phonon bands we observe PI phonon bleaching in the range of the 585-cm$^{-1}$
phonon band and a slight PI absorption below 580 cm$^{-1}$. The PI phonon
bleaching consists of two peaks at 600 and 660 cm$^{-1}$ respectively with a
dip in-between at 630 cm$^{-1}$. This two PI transmission peaks are
reproducible among different runs, while the structure of the PI absorption
below 580 cm$^{-1}$ is not, and presumably arises due to increasing
instrumental noise at the lower end of the spectral range.

In the $x=0$ FM sample there is a much weaker PI absorption centered around $%
\thicksim 3000$ cm$^{-1}$ (0.37 eV). A comparison with the TD spectrum,
which shows TD transmission below 4000 cm$^{-1}$, shows that the PI
absorption is not thermally induced. The spectra in the $x=0.1$ and $x=0.2$
samples show no significant PI signal and are flat within the noise level.
TD transmission below 2000 cm$^{-1}$ centered around 1200 cm$^{-1}$ is
observed in the $x=0.1$ sample. The absence of any PI signal in the same
sample confirms again that no thermally induced signal is present in the PI
spectra, so we can eliminate thermal effects from the discussion.

To obtain information about the recombination dynamics of the PI carriers in
the $x=0$ AFM sample, the integrated intensity of the PI absorption peak was
measured as a function of the laser fluence and is shown in Fig. 3a. It can
clearly be seen that the integrated intensity of the peak is not
proportional to the laser fluence $\Phi $, but shows a square root
dependence on the laser fluence $\frac{\Delta {\cal T}_{PI}}{{\cal T}}%
\propto \sqrt{\Phi }$.

The temperature dependence of the PI-absorption-peak integrated intensity in
the $x=0$ AFM sample is shown in Fig. 3b. The integrated intensity quickly
diminishes with increasing temperature disappearing between 80 and 100K.
There is no significant shift of the PI-absorption peak observed with
increasing temperature and no changes in the PI-spectra are observed around
Neel temperature $T_{N}$.

\section{\protect\smallskip Discussion}

When a photon of visible light is absorbed in the sample at first the
primary hole-electron pair is created. In LaMnO$_{3}$ at the incoming photon
energy 2.4 eV the hole-electron pair corresponds to a charge transfer from
the occupied O $2p$ derived bands to the unoccupied Mn $e_{g}$ and Mn $%
t_{2g} $ derived bands.\cite{ArimaTokura93} The primary electron and hole
are expected to relatively quickly relax by exciting secondary lower energy
hole-electron pairs among other low energy excitations (phonons, magnons),
since the transport gap of 0.25 eV\cite{DeTeresaDorr98} is almost ten times
smaller than the primary pair energy.

The observed laser fluence dependence of the integrated intensity of the PI
absorption peak shown in Fig. 3a indicates that the photo-excited particle
density is proportional to the square root of the laser fluence. In the
simplest model the photo-excited particle density $n_{pe}$ is governed by: 
\[
\frac{dn_{pe}}{dt}=\alpha \Phi -r\text{,} 
\]
where $\alpha $ is constant, $\Phi $ laser fluence and $r$ recombination
rate. In steady state (the laser photoexcitation is pseudo-continuous in our
experiment) $n_{pe}$ is {\em time independent} and taking ito account
experimental fact that $n_{pe}\varpropto \sqrt{\Phi }$ it follows 
\[
r\varpropto \Phi \varpropto n_{pe}^{2}\text{.} 
\]
This clearly indicates a biparticle recombination process where two
independent photoexcited particles interact during recombination. The
observed PI absorption peak therefore most likely corresponds to the
excitations of individual electron-like and/or hole-like charge carriers
created during relaxation of the primarily photoexcited electron-hole pairs.

Similar as in high $T_{c}$ cuprates, in (La$_{1-x}$Sr$_{x}$Mn)$_{1-\delta }$O%
$_{3}$ a PI signal of significant magnitude is observed only in the lower
range of hole doping.\cite{Mertelj-unpublished} Since the laser
photoexcitation and measurement are pseudo-continuous, the photoexcited
carrier lifetimes need to be quite long for any significant photoexcited
carrier density to build up. Thus the absence of the PI signal at higher
doping levels may not necessarily mean the absence of photoinduced carriers
or polarons, but is more likely that it signifies shorter PE lifetimes. As a
consequence we mainly focus on the $x=0$ samples in the rest of discussion
where the PI signal is observed.

In (LaMn)$_{1-\delta }$O$_{3}$ the majority of Mn ions have one electron in
the split $e_{g}$ orbitals surrounded by a static JT distortion of the MnO$%
_{6}$ octahedra\cite{HuangSantoro97,ElemansVanLaar71}. If the
photoexcitation results in an additional photoexcited hole in the occupied
orbital, or an additional photoexcited electron is put into the second JT
split empty $e_{g}$ orbital, the Jahn-Teller mechanism is disabled and the
JT lattice deformation is reduced around this site. The photoexcited charge
carriers can thus form {\em anti-JT polarons}, which behave very much like
ordinary polarons. One expects the characteristic shape of the PI absorption
is the same as in the case of normal polarons. Indeed we find that the shape
of the observed PI absorption peak is consistent with the theoretically
predicted absorption due to a photon assisted hopping of small polarons\cite
{Klinger63,ReikHeese67,Emin93} as seen from the analysis that follows.

In Fig. 2a a fit of absorption due to a small polaron hopping given by Emin 
\cite{Emin93} is shown for both $x=0$ samples assuming that $\alpha $ $%
\varpropto -\frac{\Delta T}{T}$\cite{KimHeeger87}: 
\[
\alpha \varpropto \frac{1}{\hbar \omega }\exp (-\frac{(2E_{b}-\hbar \omega
)^{2}}{4E_{b}\hbar \omega _{ph}}) 
\]
where $\alpha $ is the absorption coefficient, $E_{b}$ is the polaron
binding energy, $\omega $ the incoming photon frequency and $\omega _{ph}$
the polaron phonon frequency. It can be seen that the theoretical prediction
fits well to the data with the small polaron binding energies $%
E_{b}=350\,\pm 8$ meV in the $x=0$ AFM and $E_{b}=200\,\pm 10$ meV in the $%
x=0$ FM samples respectively. The polaron binding energies $E_{b}$ and
polaron phonon frequencies $\omega _{ph}$ obtained from the fit are
summarized in Table I. The obtained small polaron binding energies are
similar to those inferred from transport measurements\cite{DeTeresaDorr98}
and {\em decrease} {\em with increased hole doping} in a similar manner.

The frequencies of the polaron phonons from the fit are in the region of the
oxygen related modes as expected. Except for the phonon bleaching we do not
observe any photoinduced local modes (PILM) in our spectra. The reason might
be that the polaron phonon frequencies (indicated by Table 1) fall below the
frequency range where we can reliably measure photoinduced absorption
features .

Finally, it is instructive to compare the PI absorption spectrum with IR
conductivity spectra of chemically doped compounds to see whether the PI
carriers show any similarity to the carriers introduced by chemical means.
Indeed, Okimoto {\em et al.}\cite{OkimotoKatsufuji97} observe in the $x=0.1$
La$_{1-x}$Sr$_{x}$MnO$_{3}$ a broad absorption peaked around 0.5 eV which is
absent at room temperature and increases in intensity with decreasing
temperature. Consistent with our present interpretation they attribute it to
localization of JT polarons at low temperatures and relate the peak energy
of 0.5 eV to a JT polaron binding energy\cite{MillisShraiman96} $E_{pJT}$.
In addition, the reported absorption feature also shifts to lower energy as
doping is increased in agreement with our photoinduced measurements.\cite
{OkimotoKatsufuji97} Midgap state with a similar peak energy and similar
doping dependenence was also observed by Jung {\em et al.}\cite{JungKim98}
in La$_{1-x}$Ca$_{x}$MnO$_{3}$.

On the other hand Machida {\em et al.}\cite
{MachidaMoritomo98,MachidaMoritomo98clust} and Quijada {\em et al.}\cite
{QuijadaCerne98} report temperature dependent features in optical spectra of
manganites with peak energy in the $1-1.5$ eV range depending on cation
composition on the rare earth (R) site. The temperature dependence of
intensity and position of these peaks is consistent with small JT polaron
disappearance below T$_{C}$. The peak energies in these experiments are much
higher than $2E_{b}$ obtained from transport\cite{DeTeresaDorr98}, Raman
scattering\cite{YoonLiu98} and extrapolation of our data. Therfore they can
not be directly related to the polaron hopping from Mn$^{3+}$ to Mn$^{4+}$
ion at energy $2E_{b}$ as suggested in ref. \cite{MachidaMoritomo98} but
rather to the charge transfer between $e_{g}$ orbitals on neighboring Mn$%
^{3+}$ ions\cite{QuijadaCerne98} at energy $2E_{b}+U$. Here $U$ is the $%
e_{g}-e_{g}$ onsite Coloumb repulsion. This is additionaly suported by the
existence of two midgap peaks in La$_{1-x}$Ca$_{x}$MnO$_{3}$, one in the
range $0.2-0.8$ eV and another in the range $1.1-1.6$ eV depending on $x$
attributed to the photon assisted polaron hopping from Mn$^{3+}$to the
unoccupied neighboring Mn$^{4+}$ and charge transfer from Mn$^{3+}$ the
occupied neighboring Mn$^{3+}$ respectively.\cite{JungKim98}

\section{Conclusions}

A photoinduced midinfrared absorption peak is observed in weakly hole doped
(La$_{1-x}$Sr$_{x}$Mn)$_{1-\delta }$O$_{3}$ and is attributed to absorption
due to photon assisted hopping of anti-JT polarons. The theoretical model
for a small-polaron absorption\cite{Emin93} fits well to the experimental
data indicating that {\em the anti-JT polarons are small} with polaron
binding energies of 200-350 meV, decreasing with increasing hole doping. The
polaron-phonon frequencies are suggested to be in the 200-300 cm$^{-1}$
frequency range. We conclude by noting that the values of binding energies
inferred from our data are in good agreement with previous transport\cite
{DeTeresaDorr98} and optical\cite{JungKim98} measurements, exhibiting a
similar trend of decreasing with increasing hole doping.

\section{Acknowledgments}

We would like to thank V.V. Kabanov for fruitful discussions.

\begin{table}
\caption{The small polaron binding energy $E_{b}$ and the phonon frequency $ \omega _{ph}$ as obtained from the fit of absorption due to a small polaron
given by Emin\protect\cite{Emin93}.}
\begin{tabular}{llll}
sample &  & $E_{b}$ (meV) & $\omega _{ph}$ (cm$^{-1}$) \\ \hline
(LaMn)$_{1-\delta }$O$_{3}$ (AFM)& $T_{N}\approx 140$ K & $350\pm 8$ & $ 300\,\pm 40$ \\ 
(LaMn)$_{1-\delta }$O$_{3}$ (FM) & $T_{C}\approx 170$ K & $200\pm 10$ & $ 230\,\pm 76$ \end{tabular}
\end{table}
%

\section{Figure Captions}

Figure 1. The IR transmittance of the four samples with different Sr doping
and cation deficiency. The shapes of the spectra beyond $\sim 3000$ cm$^{-1}$
are influenced by light scattering in the pellet. The inset shows the real
part of the AC susceptibility $\,$as a function of temperature of the four
samples. The labeling is the same as in the main panel.

Figure 2. Low temperature ($T=25$K) a) PI transmittance and b) TD
transmittance spectra as a function of doping. The spectra are vertically
shifted for clarity and the thin lines represent the zero for each spectrum.
Thick lines represent the small polaron absorption\cite{Emin93} fit to the
data.

Figure 3. The integrated PI absorption intensity in the $x=0$ AFM sample as
a function of a) the laser fluence $\Phi $ and b) temperature $T$. The solid
line in the panel a) is $\sqrt{\Phi }$ fit to the data and the dashed line
in the panel b) is a guide to the eye.

\end{document}